\documentclass[prd,twocolumn]{revtex4}

\usepackage{amsmath}

\def\Io{{\openone}}

\def\Ro{{\bf R}}
\def\antisymmetric{{\overline\varepsilon}}

\def\qcharge{a}
\def\charge{b}
\def\coupling{\lambda}

\begin{document}

\title{Model of a quantum particle in spacetime}

\author{  Jan Naudts}
\email{Jan.Naudts@ua.ac.be}
\affiliation{
          Departement Natuurkunde, Universiteit Antwerpen UIA,\\
          Universiteitsplein 1, 2610 Antwerpen, Belgium
}
\author{  Maciej Kuna}
\email{maciek@mifgate.mif.pg.gda.pl}
\affiliation{
	Wydzia\l\ Fizyki Technicznej i Matematyki Stosowanej,
	Politechnika Gda\'{n}ska,\\
	ul.~Narutowicza 11/12, 80-952 Gda\'{n}sk, Poland
}
\date{v6, December 2000}

\begin{abstract}
Doplicher, Fredenhagen, and Roberts \cite{DFR94,DFR95} proposed a 
simple model of a particle in quantum spacetime. We give a new
formulation of the model and propose some small changes and additions
which improve the physical interpretation. In particular, we
show that the internal degrees of freedom $e$ and $m$ of the particle represent
external forces acting on the particle. To obtain this result we
follow a constructive approach. The model is formulated as a covariance
system. It has projective representations in which not only the
spacetime coordinates but also the conjugated momenta are
two-by-two noncommuting. These momenta are of the form $P_\mu-(b/c)A_\mu$,
where $b$ is the charge of the particle.
The electric and magnetic fields obtained from the vector potential $A_\mu$
coincide with the variables $e$ and $m$ postulated by DFR.
Similarly, the spacetime position operators are of the form
$Q_\mu-(al^2/\hbar c)\Omega_\mu$ where $a$ is a generalized charge,
$l$ a fundamental length, and with vector potentials $\Omega_\mu$
which are in some sense dual w.r.t.~the $A_\mu$.
\end{abstract}
\maketitle

%%%%%%%%%%%%%%%%%%%%%%%%%%%%%%%%%%%%%%%%%%%%%%%%%%%%%%%%%%%%%%%%%%%%%%%%%%%%%%%
\section{Introduction}

The DFR-model, introduced in 1994 by Doplicher, Fredenhagen, and Roberts \cite{DFR94,DFR95},
describes a relativistic quantum particle with internal degrees of freedom
$e$ and $m$ which behave under Lorentz transformations as
electric resp. magnetic field vectors.
It is one of the simplest models of quantum spacetime, and as
such received a lot of attention in the literature.
E.g. \cite{MM98} adopt the basic assumption of
the DFR-model that the time-position commutators $[Q_\mu,Q_\nu]$
commute with all observables.

In previous work \cite {NJ99, NK00} we have reformulated the model as a
covariance system. It is common to study a quantum system starting 
from a Lie group $X$ of relevant symmetries. In a covariance system this 
group is supplemented with an algebra of observables which transform into 
each other under the actions of the symmetry group. We expect that all models 
of quantum mechanics and quantum field theory can be described as 
covariance systems. For example, standard quantum mechanics of a 
nonrelativistic particle can be described as a covariance system 
consisting of an algebra of functions of position together with the 
Galilei group --- see \cite {NK00}.

In this covariance approach it is important to remember that
unitary representations of a symmetry group are allowed to be projective.
In particular, for the model under study, nonvanishing time-position commutators
$[Q_0,Q_\alpha]\not=0$, $\alpha=1,2,3$, and nonvanishing momentum commutators
$[P_\mu,P_\nu]\not=0$ are obtained by considering projective
representations of the group of shifts in spacetime and in momentum space.
Indeed, let $p$ and $p'$ be two shifts in momentum space,
and let $U$ denote the projective representation. Then a phase factor $\xi$
is allowed in the composition
\begin{equation}
U(p)U(p')=\xi(p,p')U(p+p')
\label{projrepr}
\end{equation}
Now, write $\xi$ into the form
\begin{equation}
\xi(p,p')=\exp\left({i\over 2}\,\sum_{\mu,\nu=0}^3p_\mu{\cal Q}_{\mu,\nu} p'_\nu\right)
\end{equation}
with $\cal Q$ an anti-symmetric matrix.
The time-position operators $Q_\mu$ are the generators
of shifts in momentum space
\begin{equation}
U(p)=\exp\left(-i\hbar^{-1}\sum_{\mu=0}^3p_\mu g_{\mu,\mu}Q_\mu\right)
\label{gen}
\end{equation}
(the metric tensor $g$ is diagonal with eigenvalues $1$, $-1$,
$-1$, $-1$).
Combination of (\ref{gen}) and (\ref{projrepr})
implies the following commutation relations
\begin{equation}
[Q_\mu,Q_\nu]=-i\hbar^2g_{\mu,\mu}{\cal Q}_{\mu,\nu}g_{\nu,\nu}
\end{equation}
In the DFR-model the r.h.s.~of the latter expression
is an operator which commutes with all other observables.
Hence it is clear that also the phase factor $\xi(p,p')$
in (\ref{projrepr}) should be allowed to be an operator.
Unitary representations with operator-valued phase factors
have been studied in \cite{NJ99}. From a physical point of view
they are acceptable if they correspond with gauge freedoms
of the model, in other words, if the wavefunctions $\psi$
and $\xi(p,p')\psi$ describe the same state of the system.
This is obviously the case if $\xi(p,p')$ commutes with all
observables.

% list of changes

Small changes of and additions to the original DFR-model are necessary to clarify the 
structure of the model. In the present paper we limit ourselves to 
the description of a single particle. In \cite{DFR94,DFR95} also fields 
are considered. The technicality of the latter makes it hard to analyze 
the field version with the same depth as is possible for the single 
particle version. The main result of the present paper is the 
identification of the internal degrees of freedom $e$ and $m$ as constant external 
fields. It suggests that the next item to study, after the one-particle 
model, is not the field version of the model, but the interaction of a 
single particle with varying and fully quantized external fields.

An important difference with DFR is that we consider not only 
noncommuting spacetime coordinates but also noncommuting momentum 
operators. This is a deliberate choice. It is made possible by 
considering representations which are also projective for shifts in 
spacetime, the generators of which are (proportional to) the momentum 
operators. The consequences of making this choice will become clear 
further on. While considering these projective representations
it turns out to be obvious to allow that the metric tensor
$g$ depends on the internal degrees of freedom $e$ and $m$.
We use the notation $\gamma(e,m)$ for this $e,m$-dependent metric tensor
while $g$ always denotes the metric tensor $[1,-1,-1,-1]$
of Minkowski space.

Another modification to the model is the interchange of the two internal 
degrees of freedom $e$ and $m$ (corrected with a factor $e\cdot m$ to 
restore time reversal symmetry). This intervention is needed to allow 
for the interpretation of the internal degrees of freedom $e$ and $m$ as 
(analogs of) electric and magnetic fields. 
Finally, the latter interpretation suggests the introduction
of a coupling constant $\coupling$ and of charges $\qcharge$ and $\charge$.

%The paper is organized as follows. ...

%%%%%%%%%%%%%%%%%%%%%%%%%%%%%%%%%%%%%%%%%%%%%%%%%%%%%%%%%%%%%%%%%%%%%%%%%%%%%%%
\section{The model}

%\section{Internal degrees of freedom}

The internal degrees of freedom consist of two vectors $e$ and $m$ in ${\bf R}^3$
satisfying $|e|=|m|$ and $e\cdot m = \pm 1$. These $e,m$-pairs are the points of
the internal configuration space $\Sigma$. It consists of two
subspaces $\Sigma_+$ and $\Sigma_-$ corresponding with the two possible
signs of the scalar product $e\cdot m$.
DFR \cite{DFR94,DFR95} give an extensive justification
of this model. For our purposes it is important that under Lorentz transformations
points of $\Sigma$ transform into themselves. These transformations are
defined as follows. Given a point $e,m$ in $\Sigma$ introduce the following
antisymmetric matrix
\begin {equation}
\epsilon(e,m)=\left(\begin{array}{cccc}
0    &e_1  &e_2  &e_3 \cr
-e_1 &0    &m_3  &-m_2\cr
-e_2 &-m_3 &0    &m_1 \cr
-e_3 &m_2  &-m_1 &0\cr
\end{array}\right)
\end {equation}
Let $\Lambda$ be a Lorentz transformation. The transformation
of $\epsilon(e,m)$ using $\Lambda$ is denoted $\epsilon(e',m')$
\begin{equation}
\epsilon(e',m')=\Lambda^{-1}\epsilon(e,m)\tilde \Lambda^{-1}
\label{stsigma}
\end{equation}
It is again an antisymmetric matrix.
It is not difficult to show that $e',m'$ is again a point of $\Sigma$.
Hence, the Lorentz transformation $\Lambda$ maps the point
$e,m$ into the point $e',m'$.
Note that (\ref{stsigma}) differs from the conventions used in
\cite {NK00}. These differences are necessary because of the swap of
meaning of $e$ and $m$.

In the DFR-paper the variables $e$ and $m$ are by definition
the entries of the 4-by-4 anti-symmetric matrix appearing in
the commutation relations
\begin{equation}
[Q_\mu,Q_\nu]=il_{\rm P}^2{\cal Q}_{\mu,\nu}
\label{qqccrdfr}
\end{equation}
($l_{\rm P}$ is Planck's length).
In our notations this means that ${\cal Q}=\epsilon(e,m)$.
Our actual result
gives $\cal Q$ proportional to $\epsilon^{-1}(e,m)$.
Note that the inverse of the matrix $\epsilon(e,m)$ is given by
\begin{equation}
\epsilon^{-1}(e,m)=-(e\cdot m)\epsilon(m,e)
\end{equation}
so that again the differences are explained by the
interchange of $e$ and $m$.

In what follows we need to integrate over $\Sigma$ in a covariant way.
To do so, we select a special point $(e_0,m_0)$ in $\Sigma$
satisfying $|e_0|=|m_0|=1$.
Then each Lorentz transformation $\Lambda$ defines a point
$(e(\Lambda),m(\Lambda))$ by the relation
\begin{equation}
\epsilon(e(\Lambda),m(\Lambda))=\Lambda^{-1}\epsilon(e_0,m_0)\tilde \Lambda^{-1}
\end{equation}
Then, by definition we take
\begin{equation}
\int_\Sigma\hbox{ d}e\hbox{ d}m\,f(e,m)\equiv
\int_{\cal L}\hbox{ d}\Lambda f(e(\Lambda),m(\Lambda))
\end{equation}
where the latter is an integration over the Lorentz group $\cal L$.
It is then obvious that the integral of $f(e,m)$ over $\Sigma$ is
by construction invariant under Lorentz transformations.

%%%%%%%%%%%%%%%%%%%%%%%%%%%%%%%%%%%%%%%%%%%%%%%%%%%%%%%%%%%%%%%%%%%%%%%%%%%%%%%
\section{A fundamental length}

Many authors have proposed that at very short
distances the coordinates of spacetime should be
discrete, or that at least Heisenberg-type of
uncertainty relations should hold for time and
position operators. The argument is that
at the scale of Planck's length
\begin{equation}
l_{\rm P}=\sqrt{G\hbar c^{-3}}
\end{equation}
the quantum nature of gravitational forces
is important and changes the structure of spacetime.
Once that one accepts the relevance of the
fundamental unit of length $l_P$ all distances
can be expressed as dimensionless numbers.
In particular, one can convert inverse lengths
to lengths. Using Planck's constant $\hbar$ one
can then convert momenta into lengths. In what
follows we will use this idea of an absolute
length $l$ to convert shifts in position $q$ into
shifts in wavevector $k$ by means of the relation $k=l^{-2}q$.
However, this formula does not behave correctly under
time reversal. In the present model we can correct
for this by multiplying with the scalar product $e\cdot m$ which 
changes sign under time reversal,
i.e.~$(e\cdot m)l^{-2}q$ behaves as a wavevector
(it transforms as a pseudo-vector).

Already in 1949 Born \cite{BM49} suggested that,
in analogy with the rest mass squared given by
\begin{equation}
c^{-2}\sum_{\mu,\nu=0}^3g_{\mu,\nu}p_\mu p_\nu,
\end{equation}
also the pseudo-distance
\begin{equation}
d(q,q')=\sum_{\mu,\nu=0}^3g_{\mu,\nu}(q-q')_\mu (q-q')_\nu
\end{equation}
could have a discrete spectrum.
He proposed to introduce a new pseudo-metric,
which in our notations reads
\begin{equation}
\sum_{\mu,\nu}g_{\mu,\nu} (q_\mu q_\nu + l^{4}k_\mu k_\nu).
\end{equation}
The group of symmetries leaving this pseudo-metric invariant
is larger than the Poincar\'e group. By requiring covariance for
this larger group extra constraints are added to the theory.
See \cite{LS97}. It is straightforward to see that our analysis
of the DFR-model can be extended to include this larger
group. However, in the present paper we restrict ourselves
to the requirement of Poincar\'e invariance.

%%%%%%%%%%%%%%%%%%%%%%%%%%%%%%%%%%%%%%%%%%%%%%%%%%%%%%%%%%%%%%%%%%%%%%%%%%%%%%%
\section{Correlation function approach}

The commutation relations (\ref{qqccrdfr}) are the basis
of the DFR-paper. Here, the starting point is a correlation
function denoted $F(f;k,q;k',q')$, with $f(e,m)$ any function of $e$ and $m$,
and with $k$, $k'$, $q$, and $q'$ 4-vectors ($k$ has the meaning of a shift
in the space of wavevectors, and $q$ of a shift in space-time).
Later on we construct a Hilbert space representation
which is such that
\begin{equation}
F(f;k,q;k',q')=\langle\psi|U(k',q')\hat f U(k,q)^\dagger|\psi\rangle
\label{correldef}
\end{equation}
holds. In this expression $\psi$ is a wavefunction,
$U(k,q)$ is a projective unitary representation of the
additive group ${\Ro^4}\times {\Ro^4}$ of shifts in spacetime and
in wavevector space, and $\hat f$
is the quantisation of the function $f(e,m)$.

The technique of constructing quantum systems starting
not from commutation relations but from correlation
functions has been developed recently in a mathematical paper
\cite{NK00}. It is a generalization of the $C^*$-algebraic
approach which requires an algebraic structure together
with correlation functions determining the state of the system.
In the new approach, the $C^*$-algebra is replaced by a group of symmetries $X$
acting on 'classical' functions, e.g.~functions of
the position of the particle, or, what we do here,
functions $f(e,m)$ of the internal degrees of freedom
$e$ and $m$. One of the advantages of the formalism is
the room it leaves for projective representations
of $X$. This point is crucial for the present paper.

We need an explicit expression of $F(f;k,q;k',q')$ in closed form. 
Typically, this kind of correlation functions,
which can be expressed in closed form,
describe coherent states 
and have a gaussian form. Our {\sl ansatz} is
\begin{eqnarray}
& &F(f;k,q;k',q')
=\int_\Sigma\hbox{ d}e\hbox{ d}m\,w(e,m)
f(e,m)\cr
&\times&\xi(k,q;k',q';e,m)
\exp\left(-\frac{1}{2\coupling}s(k,q;k',q';e,m)
\right)\cr
& &
\label{quasifree}
\end{eqnarray}
This expression has been obtained by elaborating the simpler versions
found in \cite{NJ99,NK00}. In this expression $w(e,m)$
is a density function, i.e.~$w(e,m)$ is positive and normalized
\begin{equation}
\int_\Sigma\hbox{ d}e\hbox{ d}m\,w(e,m)=1,
\end{equation}
$\xi(k,q;k',q';e,m)$ is a complex phase factor,
$\coupling$ is a coupling constant discussed later on,
and $s(k,q;k',q';e,m)$ is a real function, bilinear in
$k,q$ and $k',q'$. In order to be a correlation function
(\ref{quasifree}) should satisfy conditions of
positivity, normalization, covariance, and continuity
(see \cite{NK00}). The explicit choice of $\xi(k,q;k',q';e,m)$
and $s(k,q;k',q';e,m)$, made below, satisifies these requirements.

The phase factor $\xi(k,q;k',q';e,m)$ is written in the following
way
\begin{eqnarray}
& &\xi(k,q;k',q';e,m)
=
\exp\left(
\frac {i}{2\coupling}(e\cdot m)\,u\cdot \epsilon^{-1}(e,m)u'
\right)\cr
& &\hbox{ where }u=lk+\coupling l^{-1}\eta(e,m) q\cr
& &\hbox{ and } u'=lk'+\coupling l^{-1}\eta(e,m) q'
\label{xidef}
\end{eqnarray}
It involves the 4-by-4 matrix $\eta$ given by
\begin{equation}
\eta(e,m)=(e\cdot m)\,\epsilon(e,m)\gamma^{-1}(e,m)
\end{equation}
Its function is to transform positions
into wavevectors.
As discussed before, the factor $(e\cdot m)$
is necessary because wavevectors are pseudovectors
changing sign under time reversal.
The choice of $\eta$ has to be made in such a way that
\begin{equation}
\eta(e',m')=\Lambda^{-1}\eta(e,m)\Lambda
\end{equation}
holds for any proper Lorentz transformation $\Lambda$,
when $e',m'$ are related to $e,m$ via (\ref{stsigma}).
This condition is satisfied if the matrix $\gamma(e,m)$ transforms like $\epsilon$,
i.e.
\begin{equation}
\gamma(e',m')=\Lambda^{-1}\gamma(e,m)\tilde \Lambda^{-1}
\end{equation}
should hold. Note that $\gamma(e,m)=g$ satisfies the latter condition.
Throughout this paper one can substitute $\gamma(e,m)$ by $g$.
Note that we assume in the sequel that $\gamma(e,m)$ is a symmetric matrix.

The function $s(k,q;k',q';e,m)$ is given by
\begin{eqnarray}
& &s(k,q;k',q';e,m)=u\cdot T(e,m)u\cr
& &\hbox{ where }u=l(k-k')+\coupling l^{-1}\eta(e,m) (q-q')
\end{eqnarray}
It involves a symmetric 4-by-4 matrix $T_{\mu,\nu}(e,m)$
whose elements may depend on $e$ and $m$.
At first sight the expression
$\exp\left(-(1/2\coupling)s(k,q;k',q';e,m)
\right)$ does not look Lorentz-covariant.
It is indeed necessary to make a special 'covariant' choice of the matrix 
$T_{\mu,\nu}(e,m)$. The requirement of covariance turns out to be that
it should transform in the same way as $\epsilon(e,m)$.
This means that, if the Lorentz transformation $\Lambda$
transforms $\epsilon(e,m)$ into $\epsilon(e',m')$
then also
\begin{equation}
T(e',m')=\Lambda^{-1} T(e,m)\tilde\Lambda^{-1}
\end{equation}
holds. Assume e.g.~that $T(e,m)=(1/2){\bf I}$
(half the identity matrix) whenever the length of $e$ and $m$
is equal to 1. Next define $T(e,m)$ for arbitrary $e$ and $m$
by $T(e,m)=(1/2)\Lambda^{-1}\tilde\Lambda^{-1}$ where $\Lambda$ is any
Lorentz boost for which $\epsilon(e,m)=\Lambda^{-1} \epsilon(e_0,m_0)\tilde\Lambda^{-1}$
with $e_0$ and $m_0$ vectors of length 1.

%%%%%%%%%%%%%%%%%%%%%%%%%%%%%%%%%%%%%%%%%%%%%%%%%%%%%%%%%%%%%%%%%%%%%%%%%%%%%%%
\section{Hilbert space representation}

The correlation function (\ref{quasifree}) can be used to define
a scalar product for wavefunctions of the form $\psi(k,q,e,m)$
by the formula
\begin{eqnarray}
\langle\psi|\phi\rangle
&=&\int_\Sigma\hbox{ d}e\hbox{ d}m\,
\int_{\Ro^4}\hbox{ d}k
\int_{\Ro^4}\hbox{ d}q
\int_{\Ro^4}\hbox{ d}k'
\int_{\Ro^4}\hbox{ d}q'\cr
& &\times
\phi(k,q,e,m)\overline{\psi(k',q',e,m)}
\xi(k,q;k',q';e,m)\cr
& &\times
\exp\left(-\frac{1}{2\coupling}s(k,q;k',q';e,m)\right)
\label{scalardef}
\end{eqnarray}
This scalar product defines the Hilbert space of
wavefunctions. We cannot use the more common
representation involving square integrable
wavefunctions. Therefore one should be careful with
the meaning of $|\psi(k,q,e,m)|^2$ being the
probability density of shifts in $k$- and $q$-space
and of internal degrees of freedom $e$ and $m$.

%GNS-construction

In this Hilbert space exists a unitary representation
of shifts in $k$- and $q$-space. It is given by
\begin{eqnarray}
& &U(k,q)\psi(k',q',e,m)\cr
&=&\psi(k+k',q+q',e,m)\xi(k',q';k,q;e,m)
\label{udefqst}
\end{eqnarray}
The representation is projective. Indeed, one verifies immediately,
using (\ref{udefqst}), that
\begin{equation}
U(k,q)U(k',q')=\hat \xi(k,q;k',q')U(k+k',q+q')
\label{projshifts}
\end{equation}
We use a $\hat{\,} $ to denote multiplication operators.
So, if $f$ is a function of $k,q,e,m$ then $\hat f$
is the operator which multiplies $\psi(k,q,e,m)$
with $f(k,q,e,m)$. In particular, $\hat \xi(k',q';k'',q'')$
is the operator which multiplies $\psi(k,q,e,m)$
with $\xi(k',q';k'',q'';e,m)$.

The correlation functions $F(f;k,q;k',q')$ follow from
equations (\ref{correldef}, \ref{scalardef}) if the wavefunction $\psi$ is taken as
\begin{equation}
\psi(k,q,e,m)=\delta(k)\delta(q)\sqrt{w(e,m)}
\label{vacuum}
\end{equation}
with $\delta(k)$ and $\delta(q)$ Dirac's delta function.
Remember that the wavefunctions are {\sl not} necessarily
square integrable functions so that the choice (\ref{vacuum}) is acceptable.
On the other hand, the interpretation
of $|\psi(k,q,e,m)|^2$ as a probability density of finding the
quantum particle in the state $k,q,e,m$ is {\sl not} correct.
This will be clear from the explicit expression for position
and momentum operators as given in the next section.

%%%%%%%%%%%%%%%%%%%%%%%%%%%%%%%%%%%%%%%%%%%%%%%%%%%%%%%%%%%%%%%%%%%%%%%%%%%%%%%
\section{Position and momentum operators}

The position and momentum operators $Q_\mu$ and $P_\mu=\hbar K_\mu$
are by definition the generators
of the group of shifts in wavevector space resp.~spacetime.
Let us fix conventions in such a way that
\begin{equation}
U(k,q)=\exp(-ik\cdot\hat\gamma^{-1} Q+i\hat\gamma^{-1} q\cdot K)
\end{equation}
holds.
A quick calculation using (\ref{projshifts}) gives then a result
which can be written as
\begin{eqnarray}
Q_\mu
&=&\sum_{\nu=0}^3\hat\gamma_{\mu,\nu}i\frac{\partial\,}{\partial k_{\nu}}
+\frac{1}{2}\hat q_\mu -\frac{al^2}{\hbar c}\hat \Omega_\mu
\cr
K_\mu
&=&-\sum_{\nu=0}^3\hat\gamma_{\mu,\nu}i\frac{\partial\,}{\partial q_{\nu}}
+\frac{1}{2}\hat k_\mu-\,\frac{\charge}{\hbar c}\sigma_3\hat A_\mu
\label{explop}
\end{eqnarray}
with $\sigma_3$ the operator which multiplies the wavefunction 
$\psi(k,q,e,m)$ with $e\cdot m$,
with $\qcharge$ and $\charge$ 'charges' of the particle,
with $\Omega_\mu$ given by
\begin{eqnarray}
\Omega_\mu(k,e,m)&=& -\frac{\hbar c}{2\coupling \qcharge}
\sum_{\nu=0}^3 \eta_{\mu,\nu}^{-1}(e,m)  k_\nu
\end{eqnarray}
and with $A_\mu$ given by
\begin{eqnarray}
A_\mu(q,e,m)&=&- (e\cdot m)\,
\frac{\coupling\hbar c}{2\charge l^2}
\sum_{\nu=0}^3 \eta_{\mu,\nu}(e,m)  q_\nu
\end{eqnarray}

The quantities $A_\mu(q,e,m)$ form a vector potential.
They satisfy the rather unusual condition
\begin{eqnarray}
& &\sum_{\mu,\nu}\gamma^{-1}_{\mu,\nu}(e,m)q_\mu A_\nu(q,e,m)\cr
&=&-\frac{\coupling}{\charge}\frac{\hbar c}{2l^2}
\gamma^{-1}(e,m) q\cdot \epsilon(e,m)\gamma^{-1}(e,m) q\cr
&=&0
\label{Lgc}
\end{eqnarray}

Introduce the notations
\begin{eqnarray}
E_\alpha&=&\sum_\nu\gamma_{0,\nu}(e,m)\frac{\partial A_\alpha}{\partial q_\nu}
-\sum_\nu\gamma_{\alpha,\nu}(e,m)\frac{\partial A_0}{\partial q_\nu},\cr
& &\qquad\alpha=1,2,3
\label{edef}
\end{eqnarray}
and
\begin{equation}
B_\alpha=-\sum_{\nu=0}^3\sum_{\beta,\zeta=1}^3\antisymmetric_{\alpha,\beta,\zeta}
\gamma_{\zeta,\nu}(e,m)\frac{\partial A_\beta}{\partial q_\nu},
\qquad\alpha=1,2,3
\label{bdef}
\end{equation}
with $\antisymmetric_{\alpha,\beta,\zeta}$ the fundamental antisymmetric tensor
of dimension 3.
One calculates
\begin{eqnarray}
E_\alpha
&=&-(e\cdot m)\,\frac{\coupling}{\charge}\frac{\hbar c}{2l^2}\left(
\sum_\nu\gamma_{0,\nu}\eta_{\alpha,\nu}
-\sum_\nu\gamma_{\alpha,\nu}\eta_{0,\nu}
\right)\cr
&=&\frac{\coupling}{\charge}\frac{\hbar c}{l^2}e_\alpha
\label{eeq}
\end{eqnarray}
and
\begin{eqnarray}
B_\alpha&=&(e\cdot m)\frac{\coupling}{\charge}\frac{\hbar c}{2l^2}\sum_{\nu=0}^3
\sum_{\beta,\zeta}\antisymmetric_{\alpha,\beta,\zeta}
\gamma_{\zeta,\nu}\eta_{\beta,\nu}\cr
&=&\frac{\coupling}{\charge}\frac{\hbar c}{l^2}m_\alpha
\label{meq}
\end{eqnarray}
Assume now that $\coupling$ is the fine structure constant of electromagnetism
and that $\charge$ is the charge of the proton. They are related by
\begin{equation}
\charge^2=\coupling\hbar c
\end{equation}
Then the equations (\ref{eeq}, \ref{meq}) become
\begin{eqnarray}
E_\alpha&=&\frac{\charge}{l^2}e_\alpha\cr
B_\alpha&=&\frac{\charge}{l^2}m_\alpha
\end{eqnarray}
Note that $\charge/l^2$ is the strength of the electric field
of the proton at distance $l$. One concludes that $e_\alpha$
and $m_\alpha$ can be interpreted as a magnetic resp.~electric
field, measured in absolute units which relate to
the elementary charge $\charge$ and the intrinsic length $l$.

In analogy with (\ref{Lgc}), the $\Omega_\mu(k,e,m)$ satisfy the condition
\begin{eqnarray}
& &\sum_\mu\gamma^{-1}_{\mu,\nu}(e,m)k_\mu\Omega_\nu(k,e,m)\cr
&=&-\frac{\hbar c}{2\coupling \qcharge}
\sum_{\mu,\nu,\zeta}\gamma^{-1}_{\mu,\nu}(e,m)k_\mu
\eta^{-1}(e,m)_{\nu,\zeta}k_\zeta\cr
&=&-\frac{\hbar c}{2\coupling \qcharge}
(e\cdot m)k\cdot \epsilon^{-1}(e,m)k\cr
&=&0
\end{eqnarray}
The fields $E_\alpha$ and $B_\alpha$ can be obtained from $\Omega_\mu(k,e,m)$
by
\begin{eqnarray}
E_\alpha&=&\frac{\coupling\qcharge}{l^2 \charge}
\sum_{\beta,\zeta=1}^3\sum_{\nu=0}^3
\antisymmetric_{\alpha,\beta,\zeta}
\gamma^{-1}_{\zeta,\nu}(e,m)\frac{\partial \Omega_\nu}{\partial k_\beta}
\cr
B_\alpha&=&\frac{\coupling\qcharge}{l^2\charge}\left[
\sum_{\nu=0}^3\gamma^{-1}_{0,\nu}(e,m)\frac{\partial\Omega_\nu}{\partial k_\alpha}
-\sum_{\nu=0}^3\gamma^{-1}_{\alpha,\nu}(e,m)\frac{\partial\Omega_\nu}{\partial k_0}
\right]\cr
& &
\end{eqnarray}
($\alpha=1,2,3$).
The symmetry between these relations and (\ref{edef}, \ref{bdef}) can be understood
because the matrices
$\displaystyle \frac{\partial\Omega_\mu}{\partial k_\nu}$
and
$\displaystyle \frac{\partial A_\mu}{\partial q_\nu}$
are each others inverses (up to a constant factor). Indeed, one has
\begin{eqnarray}
\sum_{\nu=0}^3\frac{\partial A_\mu}{\partial q_\nu}\frac{\partial\Omega_\nu}{\partial k_\sigma}
&=&(e\cdot m)\frac{\hbar^2 c^2}{4\qcharge \charge l^2}\delta_{\mu,\sigma}
\end{eqnarray}
with $\delta_{\mu,\sigma}$ Kronecker's delta.

%%%%%%%%%%%%%%%%%%%%%%%%%%%%%%%%%%%%%%%%%%%%%%%%%%%%%%%%%%%%%%%%%%%%%%%%%%%%%%%
\section{Commutation relations}

From (\ref{explop}) one obtains the following commutation relations
\begin{eqnarray}
\left[Q_\mu,Q_\nu\right]
&=&-\frac{al^2}{\hbar c}\sum_{\sigma=0}^3\left[i{\partial\,\over\partial k_\sigma},
\hat\gamma_{\mu,\sigma}\hat\Omega_\nu-\hat\gamma_{\nu,\sigma}\hat\Omega_\mu
\right]\cr
&=&i\frac{l^2}{2\coupling}\sum_{\sigma=0}^3\left(
\hat\gamma_{\mu,\sigma}\hat\eta^{-1}_{\nu,\sigma}
-\hat\gamma_{\nu,\sigma}\hat\eta^{-1}_{\mu,\sigma}
\right)\cr
&=&-i\frac{l^2}{\coupling}\sigma_3(\hat\gamma\hat\epsilon^{-1}\hat\gamma)_{\mu,\nu}
\label{qqccr}
\end{eqnarray}
and
\begin{eqnarray}
\left[K_\mu,K_\nu\right]
&=&\frac{\coupling}{\hbar c}\sigma_3
\sum_{\sigma=0}^3\left[i{\partial\,\over\partial q_{\sigma}},
\hat\gamma_{\mu,\sigma}\hat A_\nu-\hat\gamma_{\nu,\sigma}\hat A_\mu
\right]\cr
&=&-i{\coupling\over 2 l^2}\sum_{\sigma=0}^3
\left(\hat\gamma_{\mu,\sigma}\hat\eta_{\nu,\sigma}
-\hat\gamma_{\nu,\sigma}\hat\eta_{\mu,\sigma}
\right)\cr
&=&i\frac{\coupling}{l^2}\sigma_3\hat\epsilon_{\mu,\nu}
\label{kkccr}
\end{eqnarray}
and
\begin{eqnarray}
\left[K_\mu,Q_\nu\right]
&=&-i\hat\gamma_{\mu,\nu}
\label{kqccr}
\end{eqnarray}
As explained before, the main difference between (\ref{qqccr}) and (\ref{qqccrdfr})
comes from the interchange of $e$ and $m$. Further differences
are the appearance in (\ref{qqccr}) of
the inverse of the coupling constant $\coupling$ and of
the generalized metric tensor $\gamma(e,m)$.
If $\gamma(e,m)\equiv g$ then the only effect
is a change of sign for the commutator between the time operator
and the position operators. The appearance of factors 
$\gamma(e,m)$ in (\ref{qqccr}) and (\ref{kqccr})
is a consequence of including $\gamma^{-1}(e,m)$ in the
definition (\ref{udefqst}) of the generators $K_\mu$ and $Q_\mu$.

Many authors, e.g.~\cite {MJ92,KA94},
have studied noncanonical commutation relations
comparable with (\ref{qqccr}, \ref{kkccr}, \ref{kqccr})
-- see e.g.~the references cited in \cite{DFR94,DFR95}.
A review of these works is out of scope of the present paper.

Note that, if one takes $\hat\Omega_\mu$ and $\hat A_\mu$
equal to zero in (\ref{explop}) then one obtains a representation
describing a particle of mass zero in the off-shell
formalism of relativistic  quantum mechanics. The noncanonical
commutation relations, which we have here, are a consequence of
the presence in (\ref{explop}) of terms containing $\hat\Omega_\mu$ resp.~$\hat A_\mu$.
Now, the procedure of replacing momenta $P_\mu$ by new momenta
$P_\mu-(\charge/c)A_\mu$ is well-known from electrodynamics. Note that
the components of the new momenta $P_\mu-(\charge/c)A_\mu$ do not necessarily
commute between themselves (this fact is well known, and was used e.g.~in \cite{MJ92}
as an argument to introduce noncommuting position operators).
Hence noncommuting momenta are quite common in quantum electrodynamics.
In the present model there is not only a substitution
of $P_\mu$ by $P_\mu-\sigma_3 (\charge/c)\hat A_\mu$ but also a substitution of
$Q_\mu$ by $Q_\mu-(\qcharge l^2/\hbar c)\hat\Omega_\mu$. The latter is responsible for
the nonvanishing time-position commutators.

%%%%%%%%%%%%%%%%%%%%%%%%%%%%%%%%%%%%%%%%%%%%%%%%%%%%%%%%%%%%%%%%%%%%%%%%%%%%%%%
\section{Poincar\'e invariance}

Shifts of the particle in spacetime are described by the unitary operators $U(0,q)$.
Indeed, one verifies that
\begin{equation}
U(0,q)Q_\mu U(0,q)^\dagger=Q_\mu+ q_\mu
\end{equation}
On the other hand is
\begin{equation}
U(0,q)K_\mu U(0,q)^\dagger=K_\mu+\frac{\charge}{l^2}\sum_\nu \hat\eta_{\mu,\nu}q_\nu
\end{equation}
Clearly, the operators $K_\mu$ are not conserved under shifts in spacetime.
This is understandable because the particle moves in external fields.
%The quantity $(\charge/l^2)\hat\eta q$ has been introduced in (\ref{xidef}).
%It is precisely equal to the shift in wavevector caused by a shift by $q$ in spacetime.

Similarly, shifts in the space of wavevectors are described by the unitary operators
$U(k,0)$. Indeed, one has
\begin{equation}
U(k,0)K_\mu U(k,0)^\dagger=K_\mu+k_\mu
\end{equation}

Next we define a unitary representation $R$ of the proper Lorentz group.
The {\sl ansatz} is
\begin{equation}
R(\Lambda)\psi(k,q,e,m)=
\psi(\Lambda^{-1} k,\Lambda^{-1}q,e',m')
\label{lorentz}
\end{equation}
with $e',m'$ related to $e,m$ by (\ref{stsigma}).
The conjugate operator $R(\Lambda)^\dagger$ is given by
\begin{equation}
R(\Lambda)^\dagger\psi(k,q,e',m')=
\psi(\Lambda k,\Lambda q,e,m)
\label{lorentzconj}
\end{equation}
It is now straightforward to verify that $R(\Lambda)$ is a unitary
representation of the proper Lorentz group.

We cannot use (\ref{lorentz}) for the whole of the Lorentz
group because time reversal must be implemented as
an anti-unitary operator \cite {SJ51}
because under time reversal $q_\mu$ goes into $-g_{\mu,\mu}q_\mu$ while
$k_\mu$ goes into $g_{\mu,\mu}k_\mu$. The operator $\Theta$
given by
\begin{equation}
\Theta\psi(k,q,e,m)
=\overline{\psi(g k,-g q,-e,m)}
\end{equation}
satisfies all requirements. It obviously satisfies $\Theta^2=\Io$.
One verifies that
\begin{equation}
(\Theta\phi,\psi)=(\Theta\psi,\phi)
\end{equation}
Finally, the parity operator $P$ is defined as an isometry
between Hilbert spaces by
\begin{equation}
P\psi(k,q,e,m)=\psi(gk,gq,-e,m)
\end{equation}
The parity-inverted scalar product is given by
\begin{eqnarray}
\langle\psi|\phi\rangle'
&=&\int_\Sigma\hbox{ d}e\hbox{ d}m\,
\int_{\Ro^4}\hbox{ d}k
\int_{\Ro^4}\hbox{ d}q
\int_{\Ro^4}\hbox{ d}k'
\int_{\Ro^4}\hbox{ d}q'\cr
& &\times
\phi(k,q,e,m)\overline{\psi(k',q',e,m)}\cr
& &\times
\xi(gk,gq;gk',gq';-e,m)\cr
& &\times
\exp\left(-\frac{1}{2\coupling}s(gk,gq;gk',gq';-e,m)\right)
\label{scalardef2}
\end{eqnarray}
It satisfies
\begin{equation}
\langle P\psi|P\phi\rangle'=\langle\psi|\phi\rangle
\end{equation}

\section{Invariants}

The position and wavevector operators $Q_\mu$ and $K_\mu$
transform as expected under proper Lorentz transformations.
From (\ref{lorentz}, \ref{lorentzconj})
and the definitions (\ref{explop}) one obtains
\begin{eqnarray}
R(\Lambda)Q_\mu R(\Lambda)^\dagger
&=&\sum_\nu\Lambda_{\mu,\nu}^{-1}Q_\nu\cr
R(\Lambda)K_\mu R(\Lambda)^\dagger
&=&\sum_\nu\Lambda_{\mu,\nu}^{-1}K_\nu
\end{eqnarray}
Note that also
\begin{equation}
R(\Lambda)\hat\gamma_{\mu,\nu} R(\Lambda)^\dagger
=\Lambda^{-1}\hat\gamma\tilde\Lambda^{-1}
\end{equation}
Introduce the squared mass operator $M^2$ by
\begin{equation}
c^2\hbar^{-2}M^2=\sum_{\mu,\nu}\hat\gamma^{-1}_{\mu,\nu}K_\mu K_\nu
\label{masssquared}
\end{equation}
Then one has obviously
\begin{equation}
R(\Lambda)M^2 R(\Lambda)^\dagger=M^2
\end{equation}
Note that $M^2$ is not necessarily invariant under shifts in spacetime.

Similarly, the squared eigentime operator
\begin{equation}
\sum_{\mu,\nu}\hat\gamma^{-1}_{\mu,\nu}Q_\mu Q_\nu
\end{equation}
is also invariant under proper Lorentz transformations.

%%%%%%%%%%%%%%%%%%%%%%%%%%%%%%%%%%%%%%%%%%%%%%%%%%%%%%%%%%%%%%%%%%%%%%%%%%%%%%%
\section{Gauge transformations}

Consider the gauge transformation
\begin{eqnarray}
A_\mu&\rightarrow& A'_\mu=A_\mu+\sum_\nu\gamma_{\mu,\nu}
\frac{\partial\chi}{\partial q_\nu}\cr
\Omega_\mu&\rightarrow&\Omega'_\mu=\Omega_\mu-(e\cdot m)\frac{2\charge^2}{\hbar l^2}
\sum_\nu\gamma_{\mu,\nu}
\frac{\partial\chi}{\partial k_\nu}
\label{gauge}
\end{eqnarray}
with $\chi$ an arbitrary function of $k,q,e,m$.
Under this transformation $E_\alpha$ and $B_\alpha$,
as given by (\ref{edef}, \ref{bdef}), are invariant.
Also the commutation relations (\ref{qqccr}, \ref{kkccr}, \ref{kqccr})
are invariant. Now let 
\begin{eqnarray}
U'(k,q)&=&\exp(-ik\cdot\hat\gamma^{-1} Q'+i\hat\gamma^{-1} q\cdot K')
\end{eqnarray}
with $Q'_\mu$ and $K'_\mu$ derived from (\ref{explop}) by
substituting $A_\mu$ by $A'_\mu$.
Then $U'$ is again a projective representation of the covariance
system. It involves the same operator valued phase factor $\xi(k,q;k',q')$
because the latter depends only on the commutation relations
(\ref{qqccr}, \ref{kkccr}, \ref{kqccr}).

Fix a positive number $\kappa$, which is the rest mass of the particle
in units $\hbar$.
Assume that $\psi(k,q,e,m)$ is a solution of the eigenvalue problem
\begin{equation}
\hbar^{-2}M^2\psi=\kappa^2\psi
\end{equation}
(we do not assume that $\psi$ is a wavefunction belonging to
the Hilbert space with scalar product (\ref{scalardef})).
Then the function $\psi'(k,q,e,m)$ given by
\begin{eqnarray}
& &\psi'(k,q,e,m)\cr
&=&\exp\left(i\frac{\charge}{\hbar c} (e\cdot m)\chi(k,q,e,m)\right)
\psi(k,q,e,m)
\end{eqnarray}
is a solution of the eigenvalue problem
\begin{equation}
\hbar^{-2}(M')^2\psi'=\kappa^2\psi'
\label{transfkg}
\end{equation}
This property is what one understands by gauge invariance
of the model. In order to check that (\ref{transfkg}) holds
let us calculate
\begin{eqnarray}
\hbar^2\kappa^2\psi'
&=&\exp\left(i\frac{\charge}{\hbar c} \sigma_3\hat\chi\right)\hbar^2\kappa^2\psi\cr
&=&\exp\left(i\frac{\charge}{\hbar c} \sigma_3\hat\chi\right)M^2\psi\cr
&=&\exp\left(i\frac{\charge}{\hbar c} \sigma_3\hat\chi\right)
M^2 \exp\left(-i\frac{\charge}{\hbar c} \sigma_3\hat\chi\right)
\psi'
\end{eqnarray}
Now use that
\begin{equation}
K_\nu\exp\left(-i\frac{\charge}{\hbar c} \sigma_3\hat\chi\right)
=\exp\left(-i\frac{\charge}{\hbar c} \sigma_3\hat\chi\right)K'_\nu
\end{equation}
to obtain 
\begin{equation}
M^2 \exp\left(-i\frac{\charge}{\hbar c} \sigma_3\hat\chi\right)
=\exp\left(-i\frac{\charge}{\hbar c} \sigma_3\hat\chi\right){M'}^2
\end{equation}
so that (\ref{transfkg}) follows.

%%%%%%%%%%%%%%%%%%%%%%%%%%%%%%%%%%%%%%%%%%%%%%%%%%%%%%%%%%%%%%%%%%%%%%%%%%%%%%%
\section{Discussion}

We have shown in this paper that the variables $e$ and $m$
appearing in the DFR-model can be explained as constant
external fields. We have swapped the role of $e$ and $m$
so that $e$ is an electric and $m$ is a magnetic field vector.
As a side effect we have also shown that the non-vanishing
time-position commutators of the model arise by substituting
the spacetime position operators $Q_\mu$ by 
$Q_\mu-(al^2/\hbar c)\Omega_\mu$, together with the well-known
substitution of momentum operators $P_\mu$ by $P_\mu-(b/c)A_\mu$.
The vector potentials $\Omega_\mu$ and $A_\mu$ are
strictly linked because the field tensors
$\displaystyle \frac{\partial\Omega_\mu}{\partial k_\nu}$
and
$\displaystyle \frac{\partial A_\mu}{\partial q_\nu}$
are each others inverses, up to a constant factor.
Obviously, these findings are of interest in a more general context
than that of this particular model. In the present model
the $k$- and $q$-dependence of $\Omega_\mu$ resp. $A_\mu$
is trivial. In a more general context, we expect more complex dependency
on $k$ and $q$. In particular, nontrivial spacetime dependence of $A_\mu$
will lead to spacetime dependence of $\Omega_\mu$.

Projective representations with operator valued phase factors
play an important role in the present paper.
A more systematic study of this kind of representations
is required.
Also other aspects of the model require further investigation.
In particular, we can make the following remarks.
\begin{itemize}
\item
Throughout the paper the metric tensor $g$ has been
replaced by an operator $\hat\gamma$ because
the mathematics allows to do so. It is not clear what such an operator-valued
metric tensor means.
\item
We did not consider spin of the particle.
Introduction of a Dirac-like equation will be discussed
in a subsequent paper.
\item
We did not consider the problem of reducibility of
the covariant representation.

\end{itemize}

%%%%%%%%%%%%%%%%%%%%%%%%%%%%%%%%%%%%%%%%%%%%%%%%%%%%%%%%%%%%%%%%%%%%%%%%%%%%%%%

\end{document}